\documentclass{article}
\usepackage[T1]{fontenc}
\usepackage[utf8]{inputenc}
\usepackage[]{ismir} 

\usepackage{graphicx}
\usepackage{amsmath,amssymb,amsfonts,url}
\usepackage{cite}
\usepackage[caption=false,font=footnotesize]{subfig}

\title{Streaming Generation for Music Accompaniment}
\multauthor
{\parbox{\linewidth}{\centering Yusong Wu~$^{1}$ \quad Mason Wang~$^{2}$ \quad Heidi Lei~$^{2}$ \quad Stephen Brade~$^{2}$ \quad Lancelot Blanchard~$^{2}$ \quad Shih-Lun Wu~$^{2}$ \quad Aaron Courville~$^{1,3}$ \quad Anna Huang~$^{2}$}}
{\parbox{\linewidth}{\centering $^1$Mila, Quebec Artificial Intelligence Institute, Université de Montréal \\ $^2$MIT -- Massachusetts Institute of Technology \hspace{1.5em} $^3$Canada CIFAR AI Chair}}

\def\authorname{Y. Wu, M. Wang, H. Lei, S. Brade, L. Blanchard, S.-L. Wu, A. Courville, and A. Huang}

\begin{document}
\maketitle

\begin{abstract}
Music generation models can produce high-fidelity coherent accompaniment given complete audio input, but are limited to editing and loop-based workflows. We study real-time audio-to-audio accompaniment: as a model hears an input audio stream (e.g., vocals), it has to also simultaneously generate in real-time a coherent accompanying stream (e.g., guitar). In this work, we propose a model design considering inevitable system delays in practical deployment with two design variables: future visibility $t_f$, the temporal offset between the start of each prediction and the latest available input, and output chunk duration $k$, the number of frames emitted per call. We train Transformer decoders across a grid of $(t_f,k)$ and show two consistent trade-offs: increasing effective $t_f$ improves coherence by reducing the recency gap, but requires faster inference to stay within the latency budget; increasing $k$ improves throughput but results in degraded accompaniment due to a reduced update rate. Finally, we observe that naive maximum-likelihood streaming training is insufficient for coherent accompaniment where future context is not available, motivating advanced anticipatory and agentic objectives for live jamming. 
\end{abstract}

%
%

\begin{figure*}[t!]
  \centering
  \includegraphics[width=\linewidth]{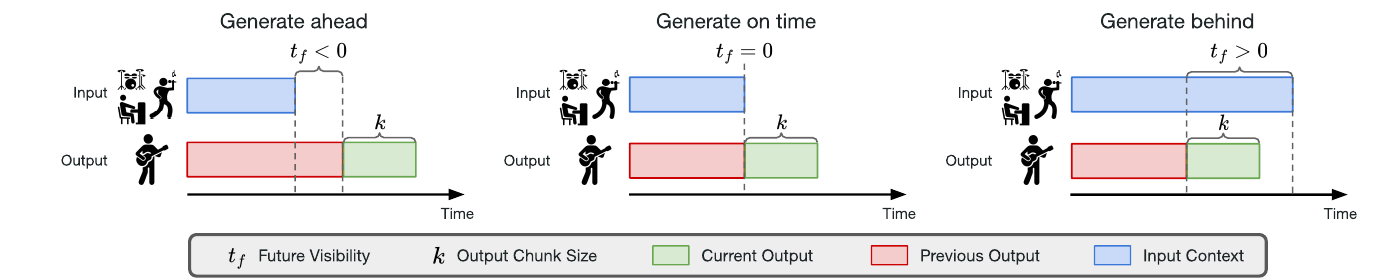}
  \caption{We formalize real-time accompaniment as a streaming, chunked prediction. At each inference step, the model predicts a length-$k$ future chunk given its previous outputs and the input history, with the future visibility $t_f$. Models with $t_f<0$ predict ahead to compensate for system latency, where the extreme case $t_f=0$ yields synchronous generation. Models with $t_f>0$ generate with input lookahead, which makes aligned playback to the performer infeasible.}
  \label{fig:generation}
\end{figure*}

\section{Introduction}
\label{sec:intro}

Live music interaction, such as jamming or accompaniment, involves reciprocal coordination, anticipation~\cite{Keller2008, roman2019delayed}, and collaborative creativity among players, giving rise to creative flow~\cite{wrigley2013experience} and social connection~\cite{trost2024live}, while remaining sensitive to communication delays~\cite{chafe2004effect}. In live acoustic accompaniment, a model must listen to an incoming audio stream and emit an aligned coherent accompanying stream in real time.

Recent multi-track generation systems leverage modern deep generative models to achieve high-fidelity accompaniment~\cite{parker2024stemgen, donahue2023singsong, rouard2025musicgen, karchkhadze2024multi, karchkhadze2025simultaneous}. However, these systems are operated offline: they require the full input context before decoding. As they typically rely on large-scale generative models, generation of audio incurs nontrivial end-to-end latency and the systems are constrained to editing and loop-based workflows rather than live interaction. In offline use, the same latency merely lengthens the wait. In live accompaniment, excessive end-to-end latency or failure to maintain real-time alignment yields misaligned, inharmonic output, even when the generated accompaniment is coherent. Several recent works target real-time accompaniment but focus on small generative models with simple symbolic outputs such as note counterpoint or chord accompaniments~\cite{wu2024adaptive, benetatos2020bachduet, wang2022songdriver}. Audio serves as a universal interface for both human perception and musical output, naturally supporting realistic timbre, effects, and expressive performance that symbolic representations cannot capture. At the same time, audio generation imposes modeling challenges, as raw waveforms are considerably longer and more low-level than symbolic sequences, requiring the model to infer musical structure and maintain coherence in both content and timbre.

In this paper, we study a framework for designing live acoustic-to-acoustic accompaniment with large generative models operating in real time. We target two deployment constraints: alignment and throughput. The accompaniment rendered and played at wall-clock time $t$ should musically align with the input audio at the same time, and generation must keep pace with playback. We formalize accompaniment as streaming chunked prediction with two design variables: the future visibility $t_f$ and the output chunk duration $k$. At step $t$, the model predicts the future chunk $y(t{:}t{+}k)$ given the input history $x(1{:}t+t_f)$ and the previous output $y(1{:}t)$. Since end-to-end latency from audio I/O, model inference, and rendering is nontrivial, a negative $t_f$ allows the model to produce audio slightly ahead to compensate, while predicting chunks of length $k$ enables buffering and parallel generation. These choices introduce trade-offs: smaller $t_f<0$ increases tolerance to latency variability but widens the recency gap and reduces reactivity; larger $k$ improves stability and throughput but lowers the update rate and delays incorporation of the latest input. This formulation separates model behavior from system constraints and exposes a latency-responsiveness design space that trade off quality.

We investigate these trade-offs using Transformer-based decoders~\cite{vaswani2017attention} trained on multi-track audio from the Slakh2100 dataset~\cite{manilow2019cutting}. We tokenize audio with a residual vector-quantized (RVQ) representation~\cite{kumar2023high} and condition the model on a random mixture of other tracks to generate one track. We sweep $t_f$ to emulate systems with different system delay budgets, and sweep $k$ to emulate different throughput and buffering regimes. We evaluate with three metrics for accompaniment quality under different system designs to show the frontier of the quality-latency-responsiveness trade-offs. We evaluate a learned coherence score and rhythm alignment between input mixture and output track, as well as audio quality on the generated track. Additionally, we conduct a listening study on selected systems, validating that our metrics align well with human perception.

Our contributions are as follows. First, we introduce a formal real-time accompaniment setup for streaming audio that identifies future visibility $t_f$ and chunk size $k$ as design axes. Second, we provide a systematic study on the Slakh2100 dataset that quantifies quality-latency-responsiveness trade-offs induced by different system designs across a grid of $(t_f, k)$. Third, we show that supervised streaming training alone is insufficient for coherent accompaniment, which motivates future work to develop training objectives that encode anticipation and coordination for live jamming. Finally, we release our codebase and pretrained checkpoints to support reproducibility and future research\footnote{Audio samples: \url{https://github.com/lukewys/stream-music-gen}}.

\section{Related Work}
\label{sec:related_work}

\textbf{Real-time Music Accompaniment Systems}\space\space
Building a computer partner that can perform with humans in real time has been a long-standing goal in computer music research~\cite{dannenberg1984line}. Early systems are predominantly non-neural. Score-following methods align a live performance to a notated score and trigger a pre-authored accompaniment in synchrony~\cite{dannenberg1984line, raphael2010music, cont2008antescofo}. Other approaches generate accompaniment with hand-crafted heuristics~\cite{lewis2003too} or by recombining phrases from a corpus that is either precompiled or built from recent input~\cite{assayag2006omax, nika2012improtek, nika2017improtek}.
Recent work has shifted toward deep generative models for accompaniment. Most systems adopt small model backbones and target simplified symbolic tasks such as note counterpoint or chord sequence prediction~\cite{benetatos2020bachduet, wang2022songdriver, jiang2020rl, wu2024adaptive, scarlatos2025realjam}. Notably, the recently released Magenta RealTime~\cite{team2025live} generates an uninterrupted stream of acoustic music and responds continuously to user input specified by weighted text prompts. In contrast, we study design choices for acoustic-to-acoustic accompaniment models that listen to a live audio stream and aim to emit music in real time.

\textbf{Real-time Control}\space\space
Real-time control has long addressed fixed and time-varying delays by predicting and scheduling outputs in advance. Dead-time compensation, typified by the Smith predictor~\cite{Smith1959}, uses an internal model to forecast the plant response over the delay and to adjust control so the commanded behavior remains time aligned. Model predictive control formalizes this lookahead through a finite prediction horizon and can incorporate known or bounded delays directly in the model or via horizon settings for robust tracking under latency~\cite{Schwenzer2021}. Recent research on real-time robot execution adopts action chunking that predicts a short sequence of future actions per inference, providing an execution buffer~\cite{black2025real}. 

\section{Streaming Accompaniment Models}

\begin{figure*}[t!]
  \centering
  \includegraphics[width=\linewidth]{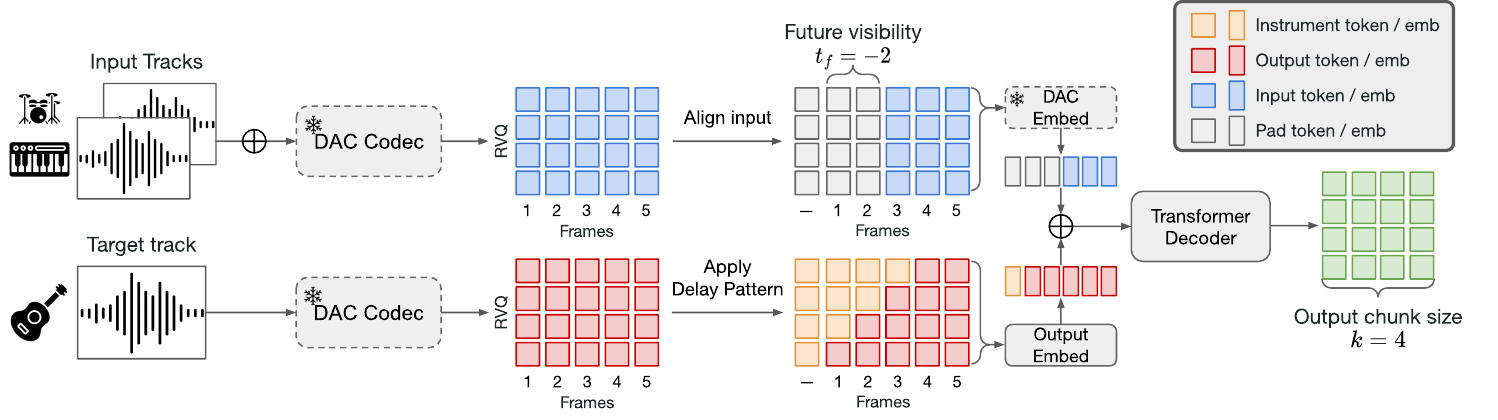}

  \caption{We implement streamed generation models as decoder-only Transformer conditioned on two streams. (Left) Audio of input tracks is mixed and both the mixture and the target stem are tokenized into RVQ codes by a DAC codec. (Middle) We apply the delay pattern to the output stream and prepend instrument token, then align the input and the output with padding to realize $t_f$. (Right) We embed input via frozen DAC codebook and output tokens with a learned embedding, before added and fed to the model, which predicts the next $k$ frames.}
  \label{fig:streaming-models}

\end{figure*}

\subsection{Problem setup}
We study streaming accompaniment with an input audio stream $x(1{:}T)$ and an output accompaniment stream $y(1{:}T)$ of equal length $T$ frames.
At the current time $t \in \{1,\dots,T\}$, the model must make musical decisions given partially observable context, i.e., it can only use the input history and its own previous output.
When deployed in a system, the rendered output at each moment should musically accompany the present input (e.g., in harmony and rhythm).
During the interaction window, the model continuously generates audio while adapting to new inputs, and ideally can anticipate near-future input so that it generates slightly in advance.
We refer to such models as \emph{streaming accompaniment models}.

We formulate one streaming step as modeling a chunk prediction with \emph{future visibility} $t_f \in [-T, T]$, the temporal offset between the start time of the prediction $t$ and the latest input; \emph{output chunk duration} $k>0$, the length of audio emitted per generation call:

\begin{equation}
\label{eq:streaming-step}
p_{\theta}\!\left(
  y\!\left(t : t{+}k\right)
  \,\middle|\,
  y\!\left(1 : t\right),\,
  x\!\left(1 : t+t_{f}\right)
\right).
\end{equation}
As shown in Figure~\ref{fig:generation}, the model with parameter $\theta$ streams the accompaniment by iteratively calling the model: to predict the output starting from $t$, it conditions on previous output $y(1{:}t)$ and input history $x(1{:}t{+}t_f)$, produces a length-$k$ chunk $y(t : t{+}k)$, then renders it. In this paper we focus on model design and training under this interface and defer interactive system engineering to future work.

\subsection{Future Visibility}
The future visibility $t_f$ intends to model the real delay from receiving the input to emitting audible output.
By allowing $t_f$ to be negative, zero, or positive, we cover three regimes: $t_f<0$ predicts ahead to compensate for system latency, $t_f=0$ is strictly synchronous, and $t_f>0$ grants lookahead to future input available in buffered or offline use. The quantity $t_f<0$ controls the \emph{recency gap} between the latest available input $x(t{+}t_f)$ and the first predicted sample $y(t)$: smaller $t_f$ increases robustness to variable delays but widens this gap.

\subsection{Output Chunk Duration}
The output chunk duration $k$ specifies the length of each generated chunk.
Setting the chunk duration $k$ creates headroom between playback and inference. While the current chunk is rendered, the next chunk can be computed, so inference calls can be scheduled within a timing window rather than at a fixed cadence. 
This asynchronous scheduling improves tolerance to compute or network jitter, enables remote serving, and in principle accommodates non-autoregressive generative models (e.g., diffusion). $k$ trades responsiveness for stability.
Smaller $k$ increases responsivenes via more frequent context updates while larger $k$ improves stability at the cost of a reduced update rate.

\subsection{Practical System Design}
The pair $(t_f,k)$ defines a model space that balances latency and reactivity. However, for real-time accompaniment models, not all $(t_f,k)$ pairs are deployable.
Small $t_f$ with small $k$ yields highly reactive behavior with frequent updates and minimal recency gap, but little tolerance to variable delays.
Moderate $k$ with a $t_f$ small enough to cover expected delays enables regular opportunities to adapt while remaining stable.

Let $\tau_{\text{sys}} > 0$ denote the end-to-end system latency from sensing input to rendering output, and let $\tau_{\text{jitter}} > 0$ denote a safety margin for variability.
A latency condition for a practical system with $t_f<0$ is $|t_f| \;\ge\; \tau_{\text{sys}} + \tau_{\text{jitter}}.$
Let $t_{\text{gen}}(k)$ be the average wall time required to generate a chunk of duration $k$, resulting in a minimum update rate of $r_{\text{update}} = 1/k$.
A throughput condition is $t_{\text{gen}}(k) \;\le\; k$, so that generation keeps pace with playback.
These conditions delimit a feasible region in the $(t_f,k)$ plane.

\subsection{Connecting to Offline Models in Unified Space}
Although practical real-time accompaniment occupies a subspace of the $(t_f,k)$ plane, Eq.~\eqref{eq:streaming-step} provides a unified formulation that also covers offline regimes with future context. Ignoring $x$ yields a decoder-only unconditional generative model on $y$. Allowing lookahead by taking $t_f>0$ grants access to future input; with full input $x(1{:}T)$ available (i.e., $t_f=T$), setting $k=1$ recovers encoder-decoder autoregressive decoding, whereas setting $k=T$ recovers non-autoregressive parallel generation models such as diffusion~\cite{karchkhadze2024multi, karchkhadze2025simultaneous} or masked language modeling (MLM)~\cite{garcia2023vampnet, parker2024stemgen}. 
Prior streaming or real-time systems instantiate specific points in the $(t_f,k)$ plane defined by Eq.~\eqref{eq:streaming-step}. Anticipatory Music Transformer~\cite{thickstun2024anticipatory} uses a fixed lookahead and corresponds to $t_f=\delta>0$, while ReaLchords~\cite{wu2024adaptive} corresponds to $t_f=0$ and $k=1$.
Thus, by sweeping $t_f$ and $k$ we interpolate smoothly from online to offline, and from single-step to chunked prediction.

\section{Experiment Setup}
\label{sec:experiment_setup}

We study latency-responsiveness trade-offs by training a grid of $(t_f,k)$ streaming models on the Slakh2100 dataset. 
We implement decoder-only Transformers operating on RVQ tokens, and we include two offline models for reference. 
We simulate streamed generation on the test set and evaluate $1024$ examples using three metrics for overall coherence, rhythmic alignment and audio quality.

\subsection{Dataset and Audio Representation}
We use the Slakh2100 dataset~\cite{manilow2019cutting}, which contains 145 hours of multi-track audio rendered from MIDI with professional virtual instruments. Each song includes a variable number of instrument stems and aligned MIDI. Following \cite{parker2024stemgen}, for each training example we randomly select one non-vocal target stem from the $N$ available stems, then form the input mixture by sampling $N_{\text{in}}\sim \mathrm{Uniform}\{1,\dots,N-1\}$ stems from the remaining set.

We extract $10\,\mathrm{s}$ windows, filtering for activity via A-weighted short-time RMS at $50\,\mathrm{Hz}$. Frames below $-60\,\mathrm{dB}$ are labeled as silent, and windows are rejected unless the input mixture and target stem share at least $50\%$ overlap of non-silent frames. We condition the model on the 18 General MIDI instrument categories~\cite{MMA_MIDI_1999}, similar to \cite{parker2024stemgen}.

We tokenize audio with a residual vector-quantized (RVQ) neural codec. Concretely, we train a Descript Audio Codec (DAC) model and use it to represent both the input mixture and the target stem as sequences of $N_q$ discrete codebook indices per frame~\cite{kumar2023high}. The DAC we train encodes and reconstructs audio at 32kHz, and produces audio codes at $50\,\mathrm{Hz}$. We train a DAC with $10$ quantizer levels but use the first four levels for modeling. To control causal dependencies for real-time use, we implement a causal variant of the codec that only encodes past context for each frame of tokens~\cite{wu-etal-2025-towards}.

\begin{figure*}[t]
  \centering
  \subfloat[COCOLA Score]{
    \includegraphics[width=0.47\textwidth]{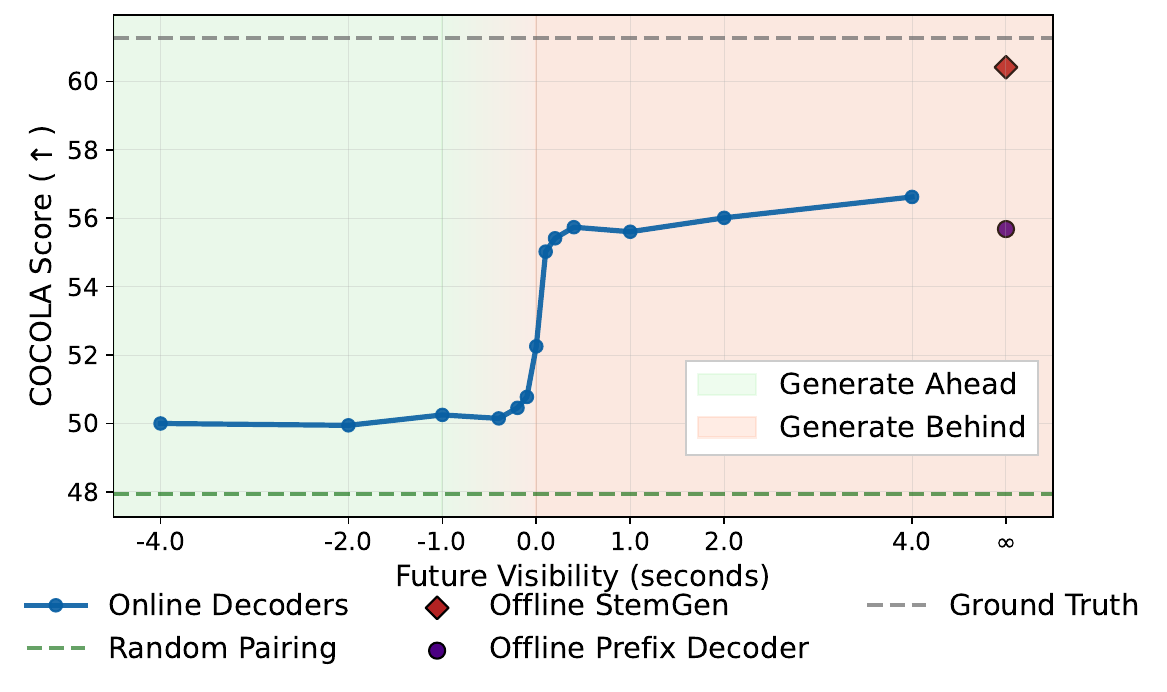}%
    \label{fig:COCOLA}}\hfill
  \subfloat[Rhythm Alignment]{%
    \includegraphics[width=0.47\textwidth]{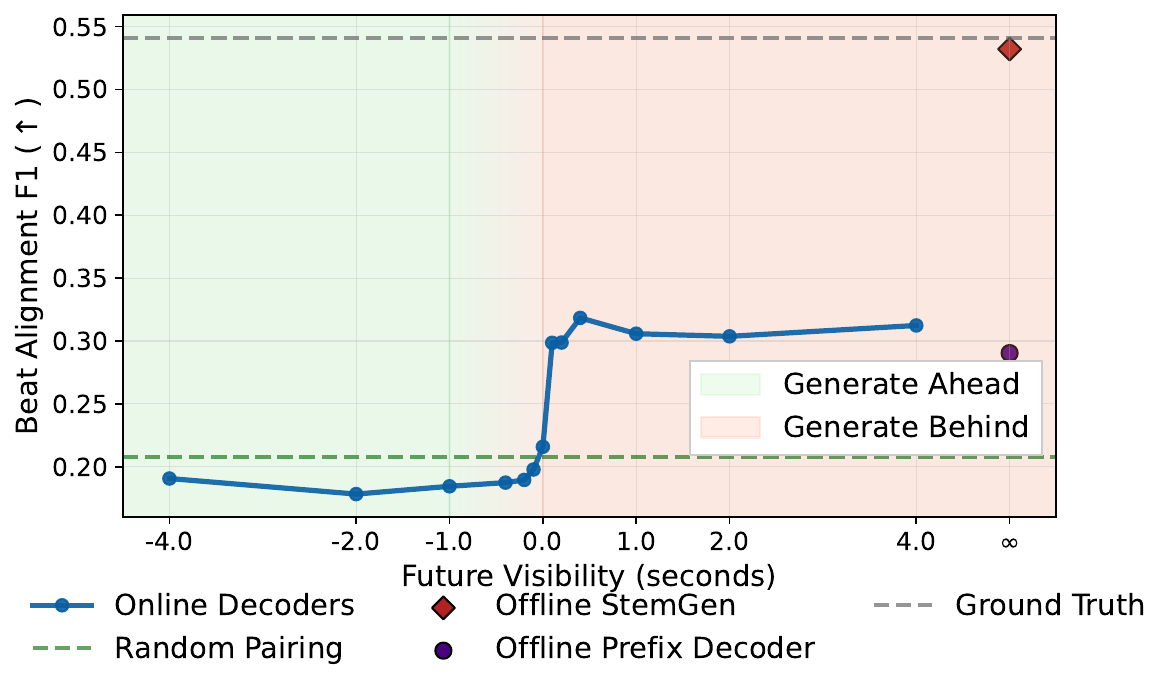}%
    \label{fig:beat}}\\ 
  
  \subfloat[Audio Quality]{%
    \includegraphics[width=0.47\textwidth]{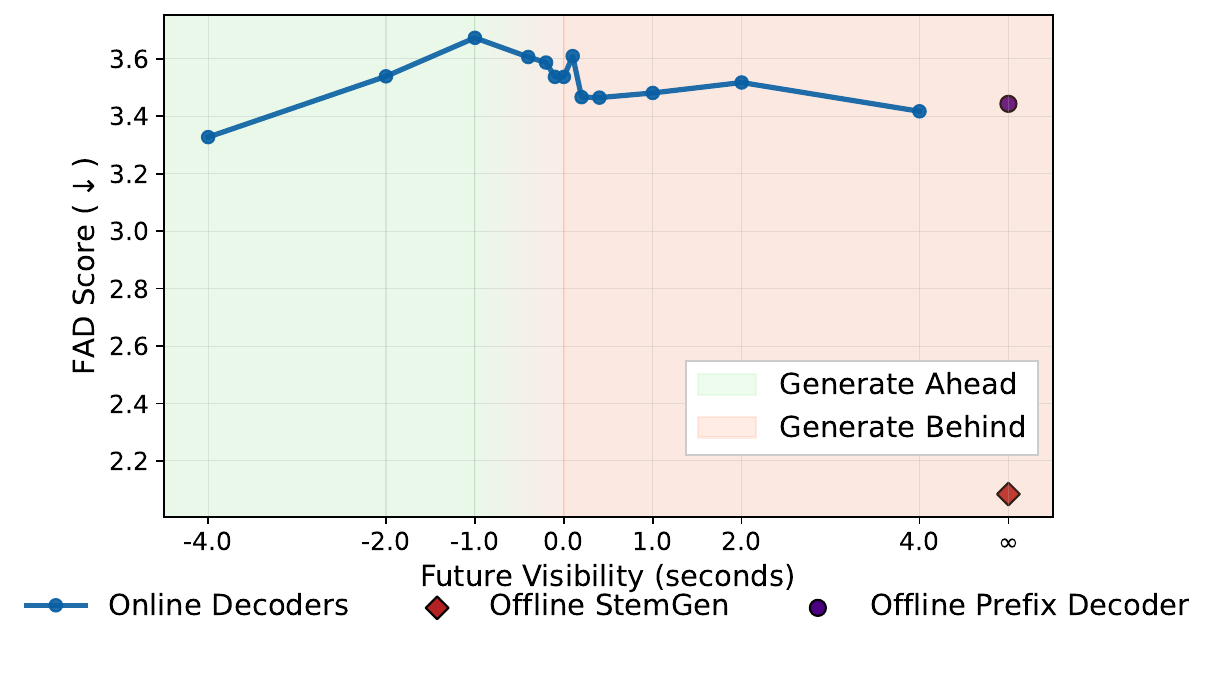}%
    \label{fig:fad}}\hfill
  \subfloat[Subjective Preference]{%
    \includegraphics[width=0.47\textwidth]{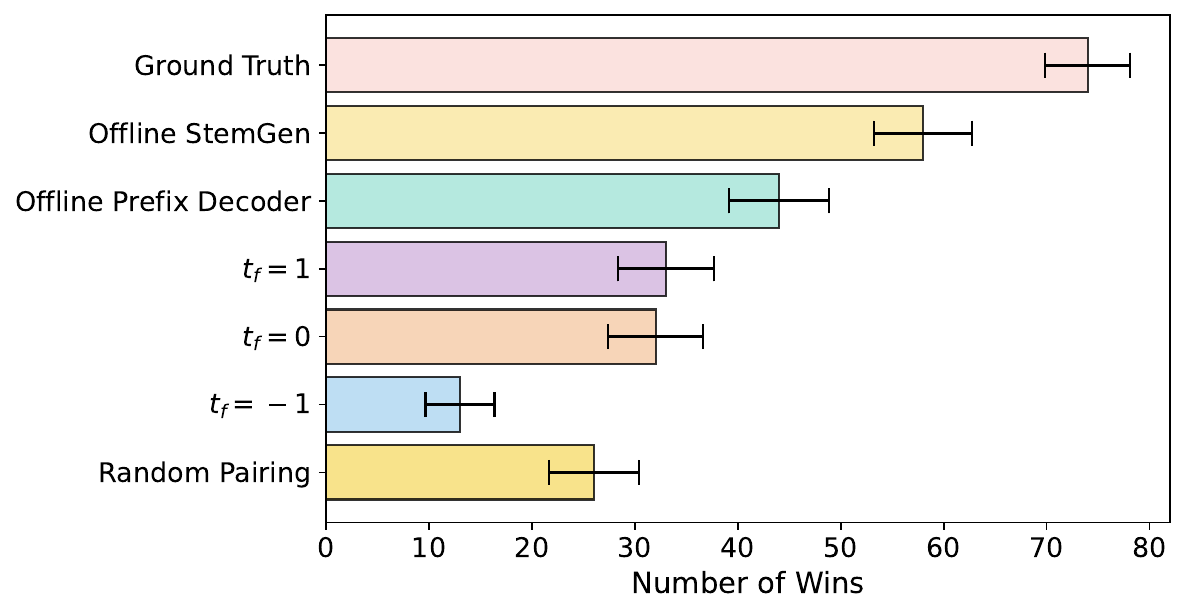}%
    \label{fig:listening}}
    
  \caption{COCOLA score, rhythm alignment, audio quality and subjective preference of generated accompaniment for streaming models with $k=1$ across different $t_f$.}
  \label{fig:metrics_row}
\end{figure*}

\subsection{Model Architecture}
We implement Eq.~\eqref{eq:streaming-step} via a decoder-only Transformer~\cite{vaswani2017attention} (Fig.~\ref{fig:streaming-models}). The backbone is a LLaMA-style~\cite{grattafiori2024llama} Transformer with 16 layers, 16 heads, and hidden size 1024. For output tokenization across RVQ levels, we adopt the ``delay pattern'' from MusicGen~\cite{copet2024simple}, which delays higher RVQ levels by one step relative to the coarse level. We choose the delay pattern over flattening RVQ levels to minimize latency: with $N_q=4$ at $50\,\mathrm{Hz}$, the delay pattern adds only $0.06\,\mathrm{s}$, whereas flattening inflates computation by a factor of $N_q$.

Let $\tau\in\{1,\dots,T_{dac}\}$ index codec frames.
Let $\mathbf{e}^x_\tau\in\mathbb{R}^{d_{\text{dac}}}$ and $\mathbf{e}^y_\tau\in\mathbb{R}^{d}$ denote the per-frame embeddings for the input mixture and the target output, respectively.
We obtain $\mathbf{e}^x_\tau$ by summing the frozen DAC codebook embeddings across RVQ levels for frame $\tau$, and we obtain $\mathbf{e}^y_\tau$ by summing a learned output codebook's embeddings across RVQ levels using the delay interleaving pattern:
\begin{equation}
\label{eq:rvq-sum}
\mathbf{e}^x_\tau \;=\; \sum_{\ell=1}^{N_q}\mathrm{Emb}^{\text{dac}}_{\ell}\!\big(c^{x}_{\tau,\ell}\big),
\qquad
\mathbf{e}^y_\tau \;=\; \sum_{\ell=1}^{N_q}\mathrm{Emb}^{\text{out}}_{\ell}\!\big(c^{y}_{\tau,\ell}\big).
\end{equation}
Here $c^{x}_{\tau,\ell}$ and $c^{y}_{\tau,\ell}$ are the RVQ code indices at level $\ell$ and frame $\tau$ for the input and output streams, $\mathrm{Emb}^{\text{dac}}_{\ell}(\cdot)\in\mathbb{R}^{d_{\text{dac}}}$ are frozen embeddings from the trained DAC, and $\mathrm{Emb}^{\text{out}}_{\ell}(\cdot)\in\mathbb{R}^{d}$ are learned embeddings used only for the output stream.

We project the input embedding to the model dimension and fuse the streams with an additive gate that encourages the model to use the input rather than relying solely on its own history:
\begin{equation}
\label{eq:fusion}
\mathbf{z}_\tau \;=\; \mathrm{LN}\!\big(\mathbf{W}\,\mathrm{LN}(\mathbf{e}^{x}_\tau)\big)\;+\;\mathbf{g}\,\odot\,\mathrm{LN}(\mathbf{e}^{y}_\tau),
\end{equation}
where $\mathrm{LN}(\cdot)$ is layer normalization, $\mathbf{W}\in\mathbb{R}^{d\times d_{\text{dac}}}$ projects $\mathbf{e}^{x}_\tau$ to the model dimension $d$, and $\odot$ denotes elementwise product.
The vector $\mathbf{g}\in\mathbb{R}^{d}$ is a learned elementwise gate initialized to zeros to bias the model toward conditioning on the input stream.

We implement instrument conditioning with a learned lookup embedding placed at the beginning of the output sequence.
For $t_f\le 0$ we pad the input stream with a PAD embedding, and for $t_f > 0$ we pad the output stream with a PAD token so that Eq.~\eqref{eq:streaming-step} is respected.

For $k=1$, we train a standard autoregressive decoder with a causal mask. For the streaming setting with chunk size $k>1$, we train a prefix decoder. Specifically, for each minibatch, we sample a prefix length $\ell$ uniformly from $\{0,\,k,\,2k,\,\ldots,\,T-k\}$. We construct each example by aligning inputs and outputs up to step $\ell$, then require the model to predict the next $k$ steps $(\ell+1,\,\ldots,\,\ell+k)$ under a causal attention mask. The loss is computed only on these $k$ target steps. This exposes the model to various prefix boundaries, enabling effective non-overlapping chunked autoregressive decoding during inference. Note that within each chunk, our model decodes autoregressively; exploring parallel intra-chunk generation (e.g., via diffusion) is left to future work.

\subsection{Model Training and Sampling}
We train two offline models for reference. A prefix decoder baseline concatenates the input and output in time and predicts only output tokens; under our abstraction this corresponds to $t_f=T$ and $k=1$. A masked language model baseline that follows StemGen over RVQ codes~\cite{parker2024stemgen} corresponds to $t_f=T$ and $k=T$. The prefix decoder matches the streaming model size, while the masked model uses 20 layers with the same heads and hidden size, following the original StemGen architecture~\cite{parker2024stemgen}.

For all prefix-decoder models, we do not use a bidirectional mask on the prefix. Instead, we apply a single causal mask over the entire sequence.
For the StemGen masked language model, we found a VampNet-style~\cite{garcia2023vampnet} confidence ranking to outperform the original StemGen ranking. At each sampling iteration per RVQ level, we compute a confidence score for token $\hat{y}_t$ as $\mathrm{conf}(\hat{y}_t)=\log p(\hat{y}_t) + temp \cdot g_t$, where $g_t\sim\mathrm{Gumbel}(0,1)$ and $temp$ is linearly annealed to $0$.

We train all models using cross entropy loss with the Adam~\cite{kingma2015adam} optimizer and a learning rate schedule with linear warmup and cosine decay (peak $1\!\times\!10^{-4}$). Streaming models and the offline prefix decoder baseline use a batch size 64; the StemGen model uses a batch size 96.
For inference, we sample decoder-only Transformers with softmax temperature $1.0$ and top-$k=200$. For the StemGen model, we use per-level maximum noise temperatures $[8.0,\,8.0,\,4.0,\,4.0]$ and $[128,\,64,\,32,\,32]$ sampling steps per level with classifier-free guidance scale $2.0$.

\subsection{Evaluation}

We assess accompaniment with three complementary objective metrics computed over the $1024$ generated test clips. We report: 
1) \textbf{Overall Coherence}: We use the pretrained COCOLA encoder~\cite{ciranni2025cocola} to compute a coherence score between the input mixture and the generated stem. COCOLA is trained in a self-supervised contrastive manner to capture harmonic and rhythmic coherence across musical audio.
2) \textbf{Rhythmic Alignment}: We estimate beats on both signals with a beat estimation model~\cite{foscarin2024beat} and report the Beat $F_1$ score using Madmom~\cite{madmom} with default settings.
3) \textbf{Audio Quality}: We report Fr\'echet Audio Distance (FAD) computed on VGGish embeddings using a reference distribution built from the test split~\cite{kilgour2018fr}.

To validate our objective metrics, we conducted a listening study with 24 participants. Participants blindly evaluated pairs of accompaniments from ground truth, random pairing, and selected models for a given input and indicated their preference. Each sample was 10 seconds long, consistent with the objective evaluation. To ease cognitive load, we selected samples with six or fewer input tracks. Significance was evaluated using a Wilcoxon signed-rank test with Bonferroni correction. Appendix~\ref{sec:fullmix} reports all results under an earlier full-mixture data setting, which the listening study uses.

For all evaluations, we use a fixed loudness pipeline: we loudness-normalize the predicted track and the target track to $-18\,\mathrm{dB}$. For the mix, input tracks are summed at equal loudness, and the predicted (or target) track is mixed $+5\,\mathrm{dB}$ relative to each input track.

\begin{figure*}[t!]
  \centering
  \subfloat[COCOLA Score Across Prompt Length]{
    \includegraphics[width=0.45\textwidth]{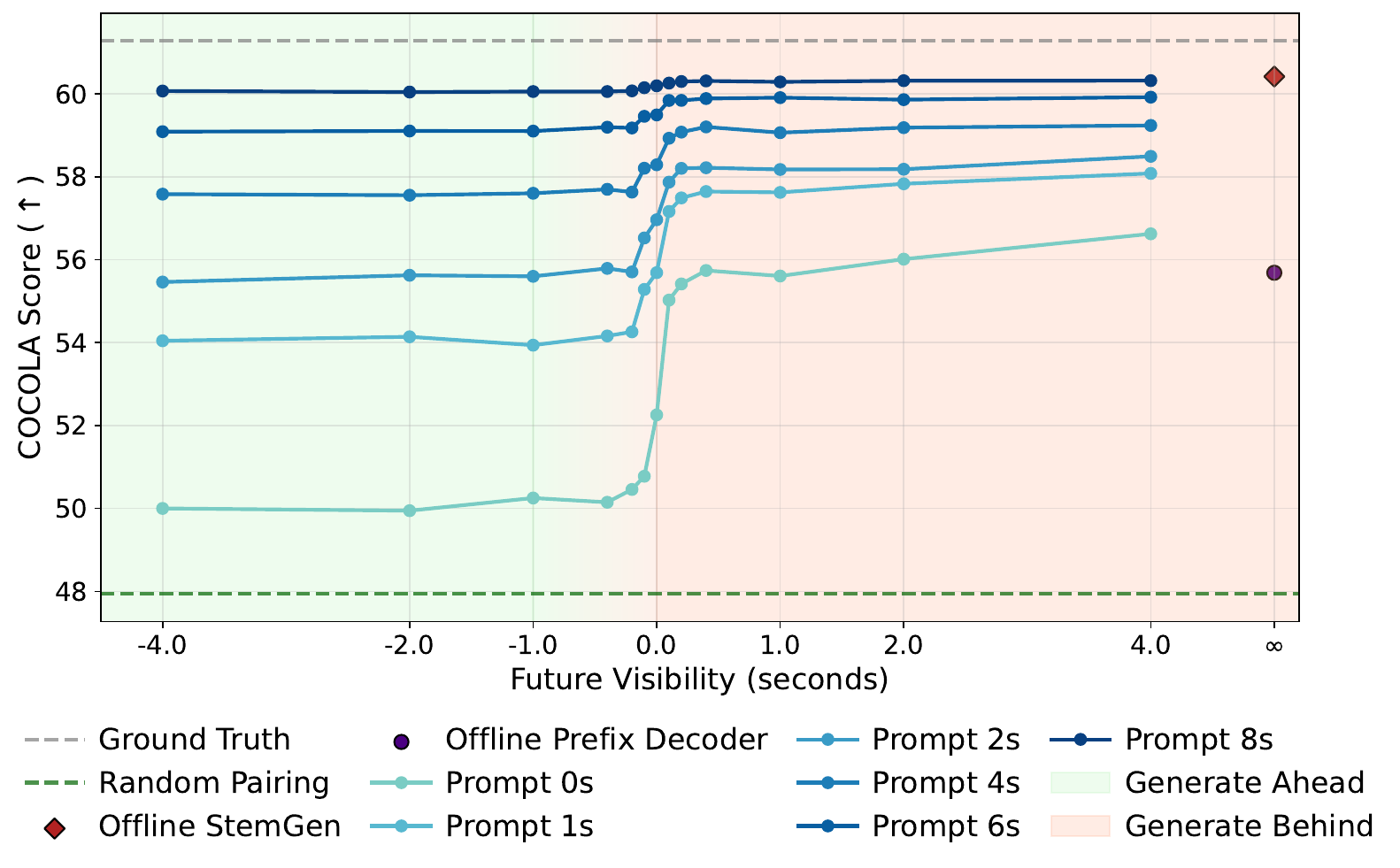}
    \label{fig:cocola_both_pred_prompt_length_normal}}\hfill
  \subfloat[COCOLA Score Compared with Random Input]{%
    \includegraphics[width=0.45\textwidth]{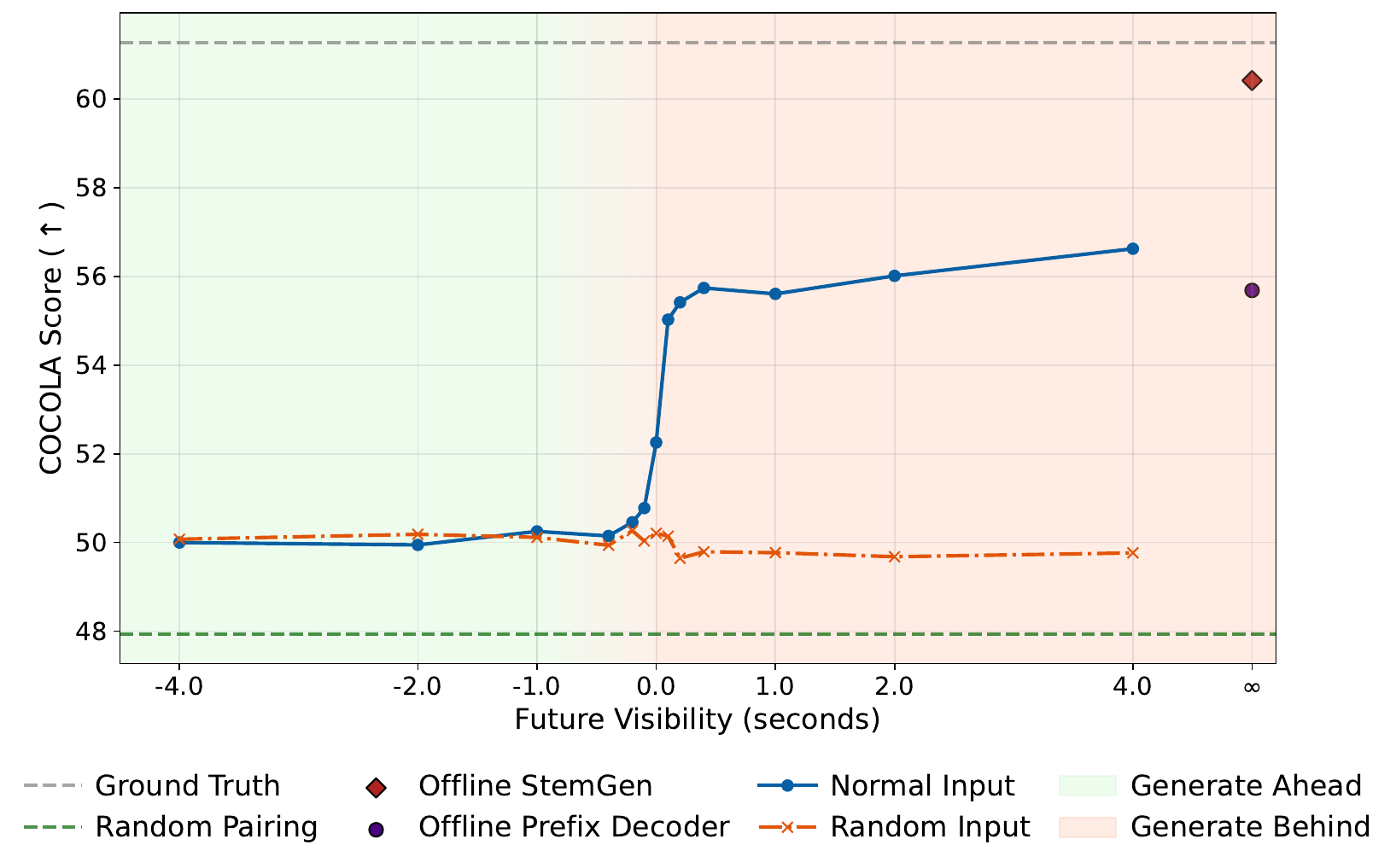}\label{fig:cocola_both_pred_prompt_0s_only}}
  \caption{Accompaniment performance for streaming models with $k=1$ evaluated across different future visibility $t_f$. Left: performance under different prompt length. Right: performance when conditioning on the paired input compared with conditioning on a random input.}
  \label{fig:inference_condition}
\end{figure*}

\begin{figure}[t!]
  \centering
  \includegraphics[width=\linewidth]{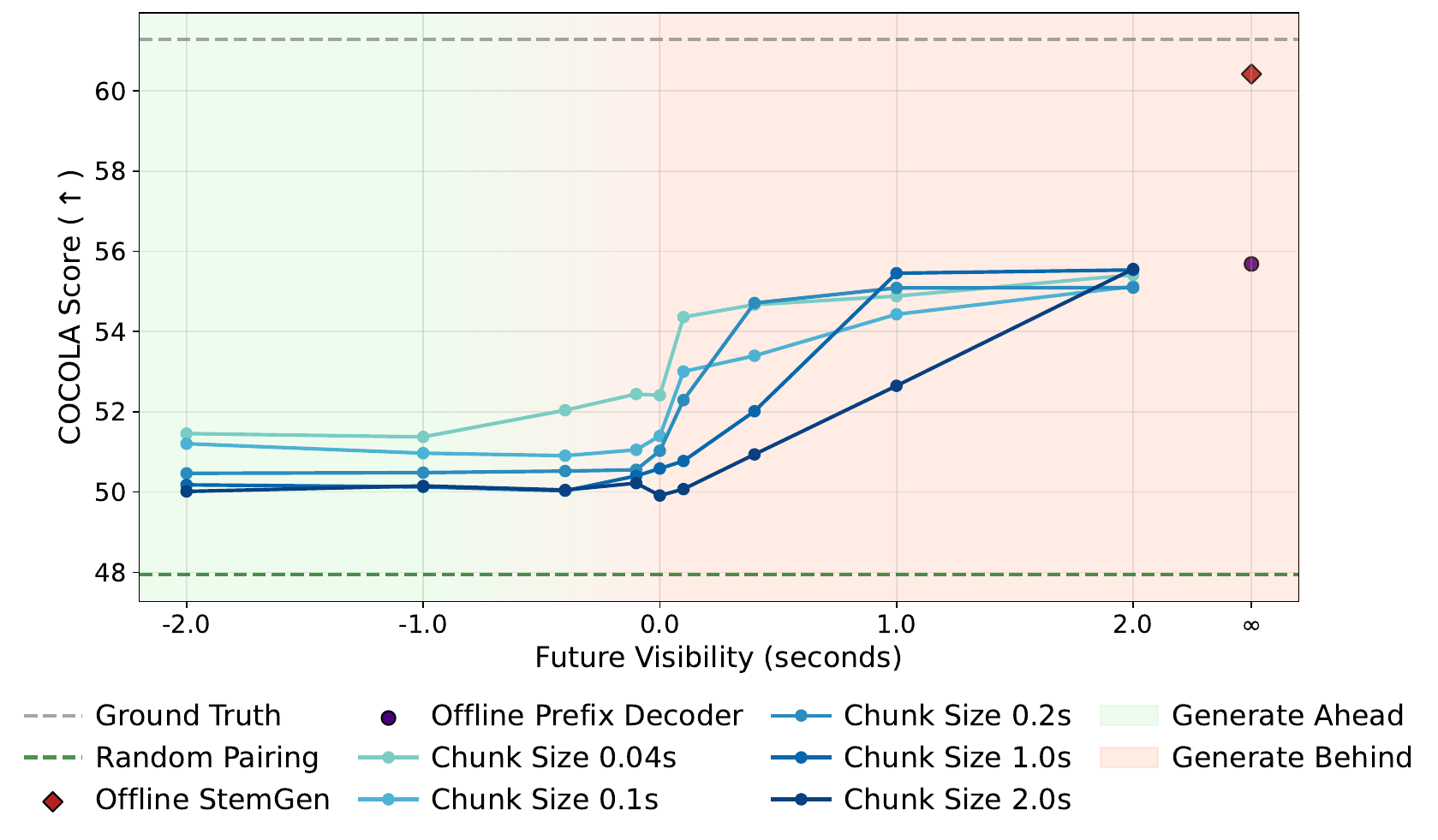}
  \caption{COCOLA score across $k>1$ and $t_f$.}
  \label{fig:COCOLA_chunk}
\end{figure}

\section{Results}
\label{sec:results}

In the following experiments, we treat $(t_f, k)$ as independent system design variables that impose latency and reactivity constraints. Our three objective metrics then measure how these constraints affect the resulting accompaniment quality: overall coherence captures harmonic and rhythmic alignment jointly, rhythmic alignment isolates temporal coordination, and audio quality assesses signal fidelity independently of the input.

We first fix $k{=}1$ frame ($0.02\,\mathrm{s}$ at $50\,\mathrm{Hz}$) and sweep $t_f\in\{\pm4,\pm2,\pm1,\pm0.4,\pm0.2,\pm0.1,0\}$ seconds.
Coherence (COCOLA, Fig.~\ref{fig:COCOLA}) and rhythmic alignment (Beat $F_1$, Fig.~\ref{fig:beat}) both show a sharp improvement when moving from $t_f\le 0$ to small positive $t_f$, with most of the gain between $t_f{=}0$ and $t_f{=}0.1\,\mathrm{s}$.
For $t_f\geq 1\,\mathrm{s}$ or $t_f\leq -1\,\mathrm{s}$ the curves largely saturate.
The offline StemGen baseline at $t_f{=}T$ outperforms all streaming models, while the offline prefix decoder is weaker but competitive.
Streaming models with small positive $t_f$ approach, and sometimes slightly exceed, the offline prefix decoder, whereas $t_f\le 0$ lies near the random-pairing floor on both metrics.

To understand the model's behavior under different visibility, we conduct two additional analyses shown in Fig.~\ref{fig:inference_condition}.
First, we investigate the model's reliance on input context by pairing each target stem with a randomly chosen input from the test set (Fig.~\ref{fig:cocola_both_pred_prompt_0s_only}). The scores with true and random inputs are indistinguishable when $t_f \le -0.2$ and differ only marginally at $t_f{=}-0.1$, which indicates that the decoder relies mainly on its own history and the instrument token rather than the input audio in negative-visibility regimes. However, as $t_f$ approaches $0$ and becomes positive, the gap increases significantly, implying that the model begins to effectively exploit the input stream for harmonic and rhythmic cues.
Second, we warm start decoding by providing a ground truth prefix of both input and output with duration $L$ seconds (Fig.~\ref{fig:cocola_both_pred_prompt_length_normal}). We observe that gains from longer prompts are largest when $t_f \le 0$, and decrease as $t_f$ grows. This suggests that in low future-visibility regimes, a short history reduces exposure bias and stabilizes local decisions, whereas with more lookahead the model already observes sufficient recent context and benefits less from a longer prefix.

The preference ordering on a selection of models (Fig.~\ref{fig:listening}) largely mirrors the objective metrics: offline models are preferred over streaming models, with larger $t_f$ preferred more. 
Notably, the streaming model with $t_f = -1$ is even less preferred than the random pairing baseline. This suggests that naive supervised training for latency compensation yields unstable generation that is perceptually detrimental.
Audio quality (FAD, Fig.~\ref{fig:fad}) varies only mildly with $t_f$ and is best at the largest $|t_f|$, while the offline StemGen achieves the best FAD.

We next vary $k$ corresponding to $\{0.04, 0.1, 0.2, 1, 2\}$ seconds across $t_f \in \{\pm2, \pm1, \pm0.4, \pm0.1, 0\}$ seconds and evaluate overall coherence (Fig.~\ref{fig:COCOLA_chunk}).
Within each $k$, the $t_f$ trend mirrors the $k{=}1$ case.
Across $k$, larger chunks hurt when $t_f<0$ because a larger fraction of each chunk lies beyond the visible input, while for $t_f>0$ chunk size matters little once $t_f\ge k$.
In particular, for $k>t_f>0$, the trailing portion $k-t_f$ is effectively unconditioned on recent input, which degrades coherence.




\section{Conclusions and Future Implications}
\label{sec:conclusion}

We formalize real-time acoustic accompaniment as streaming chunked prediction parameterized by future visibility $t_f$ and chunk duration $k$, evaluating a grid of supervised models. Both objective metrics and our listening study demonstrate that accompaniment quality improves with extended input context and more frequent updates. However, practical deployment necessitates $t_f<0$ to offset system latency, severely degrading generation quality. Consistent with prior observations, standard maximum-likelihood training on well-composed datasets proves insufficient, as these corpora lack the errors, corrective maneuvers, and co-adaptive behaviors intrinsic to live human performance. Consequently, these findings motivate the development of training objectives that explicitly encode anticipation and coordination, establishing a foundation for real-time audio accompaniment research.

Several directions follow from these findings. First, injecting synthetic imperfections (e.g., misalignments from simulated delays) during training could teach models to correct errors rather than merely predict optimal continuations. Second, while our current setup models a one-directional relationship, future systems must capture mutual co-adaptation between players. Finally, encoding distinct interactive styles (from leading to following) into generative models remains an open challenge.

\section{AI Usage Statement}
Large language models were used for grammatical editing and proofreading of the manuscript, as well as for assisting in the design and coding of the supplementary website.

\bibliography{refs}

\clearpage

\appendix
\onecolumn
\renewcommand{\topfraction}{0.95}
\renewcommand{\bottomfraction}{0.9}
\renewcommand{\textfraction}{0.05}
\renewcommand{\floatpagefraction}{0.85}
\setcounter{topnumber}{4}
\setcounter{bottomnumber}{4}
\setcounter{totalnumber}{8}
\makeatletter
\setlength{\@fptop}{0pt}
\setlength{\@fpsep}{12pt}
\setlength{\@fpbot}{0pt plus 1fil}
\makeatother

\section{Appendix}

\subsection{Acknowledgment}
We would like to thank Ke Chen for the discussion on this project.

\subsection{Additional Results}

\begin{figure*}[h]
  \centering
  \subfloat[COCOLA Harmonic]{%
    \includegraphics[width=0.49\textwidth]{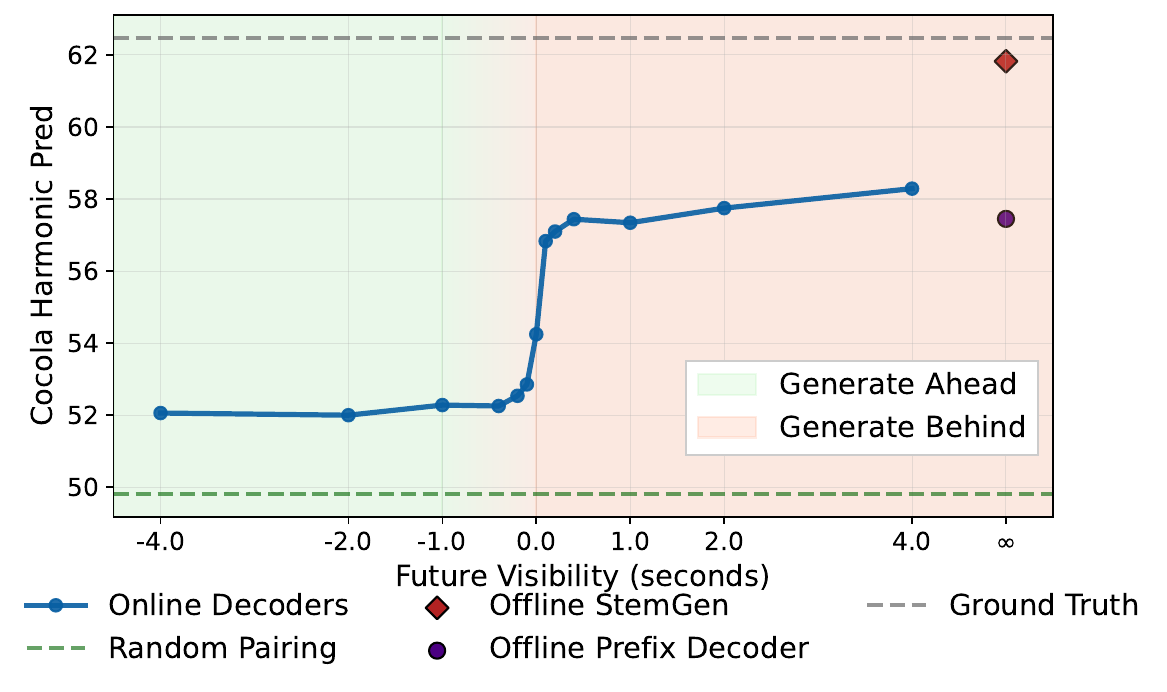}
    \label{fig:cocola_harmonic_pred}}\hfill
  \subfloat[COCOLA Percussive]{%
    \includegraphics[width=0.49\textwidth]{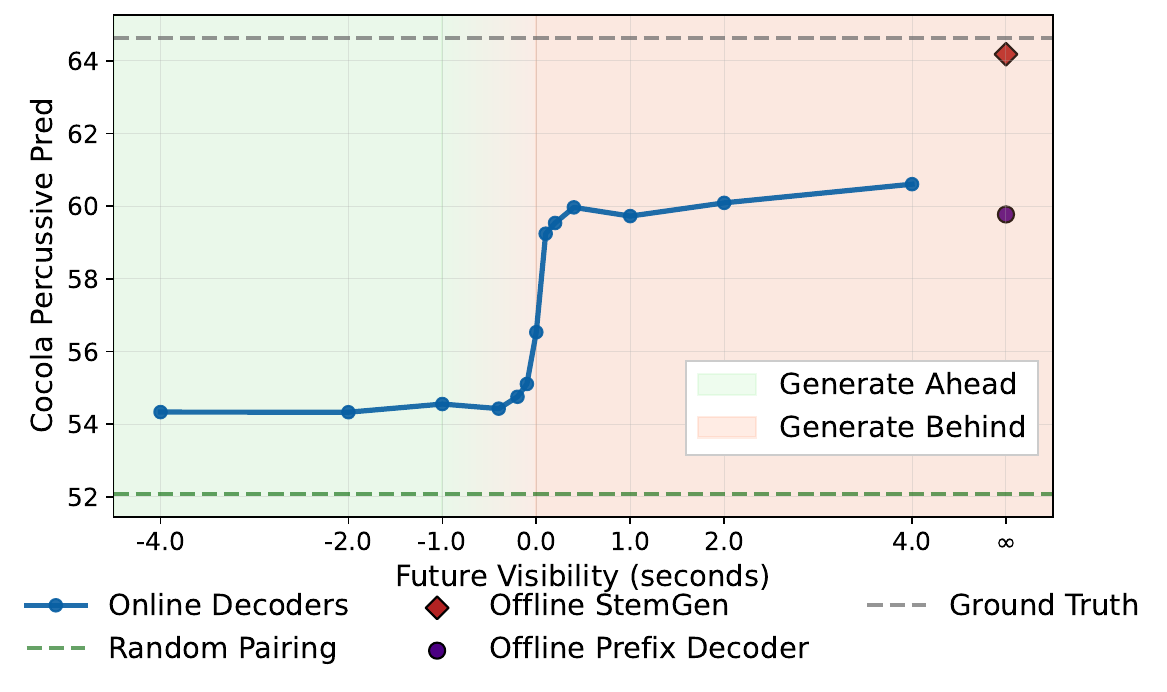}\label{fig:cocola_percussive_pred}}
  \caption{COCOLA harmonic score and COCOLA percussive score of generated accompaniment for streaming models with $k=1$ across different $t_f$.}
  \label{fig:metrics_appendix}
\end{figure*}

\begin{figure*}[h]
  \centering
  \subfloat[COCOLA Harmonic]{%
    \includegraphics[width=0.49\textwidth]{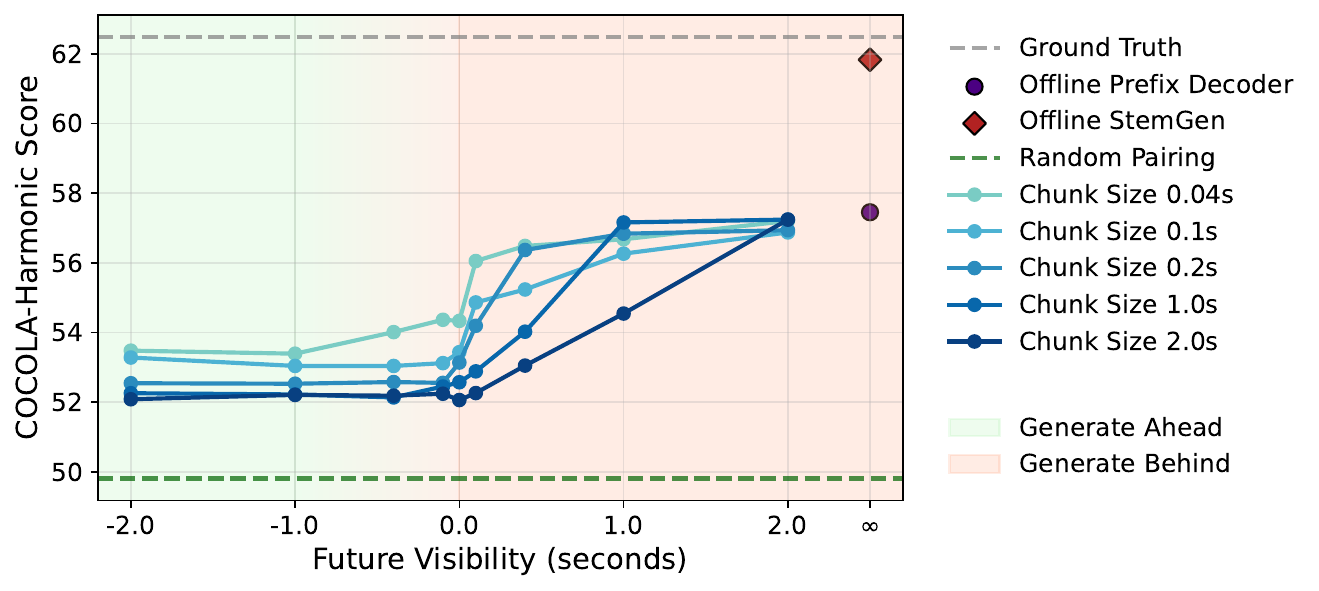}
    \label{fig:cocola_harmonic_pred_chunk}}\hfill
  \subfloat[COCOLA Percussive]{%
    \includegraphics[width=0.49\textwidth]{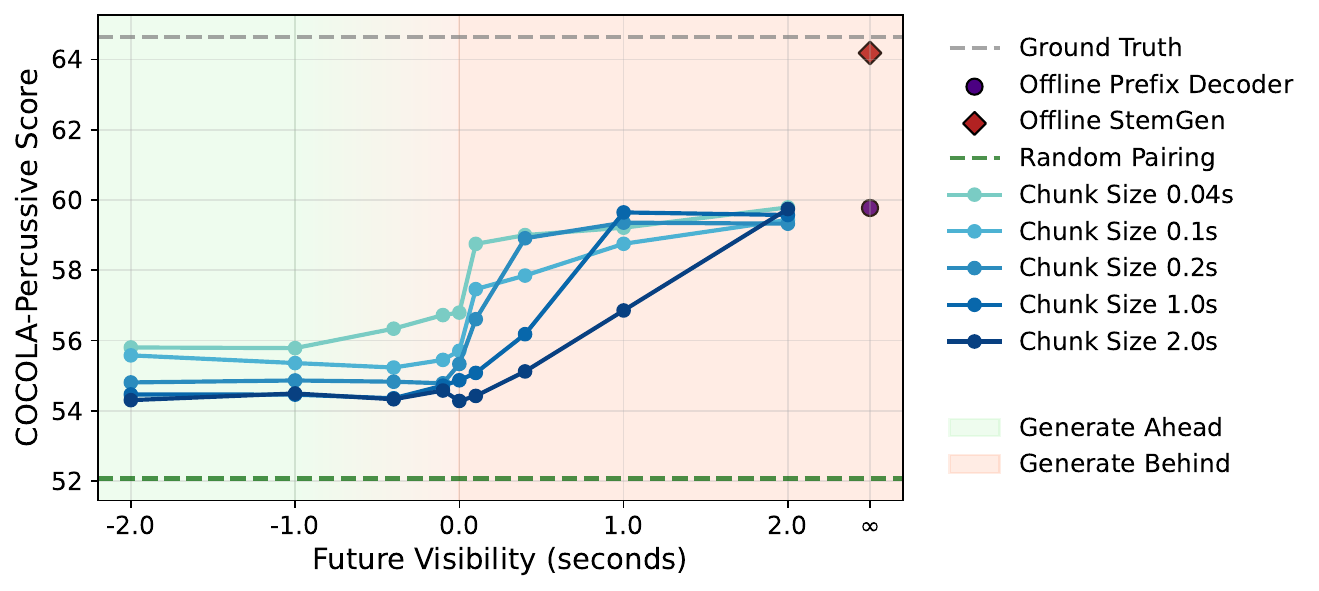}\label{fig:cocola_percussive_pred_chunk}}
  \caption{COCOLA harmonic score and COCOLA percussive score of generated accompaniment for streaming models with combinations of $k>1$ and $t_f$.}
  \label{fig:metrics_appendix_cocola_chunk}
\end{figure*}

\begin{figure*}[h]
  \centering
  \subfloat[Beat Alignment]{%
    \includegraphics[width=0.49\textwidth]{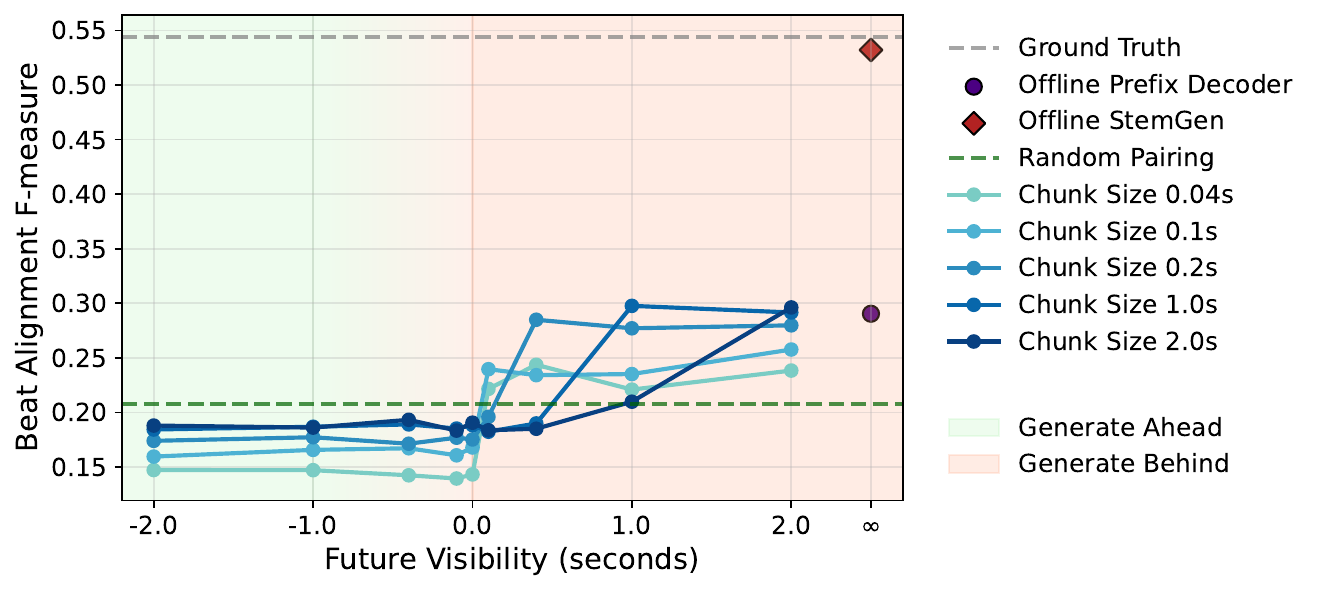}
    \label{fig:beat_align_pred_chunk}}\hfill
  \subfloat[FAD]{%
    \includegraphics[width=0.49\textwidth]{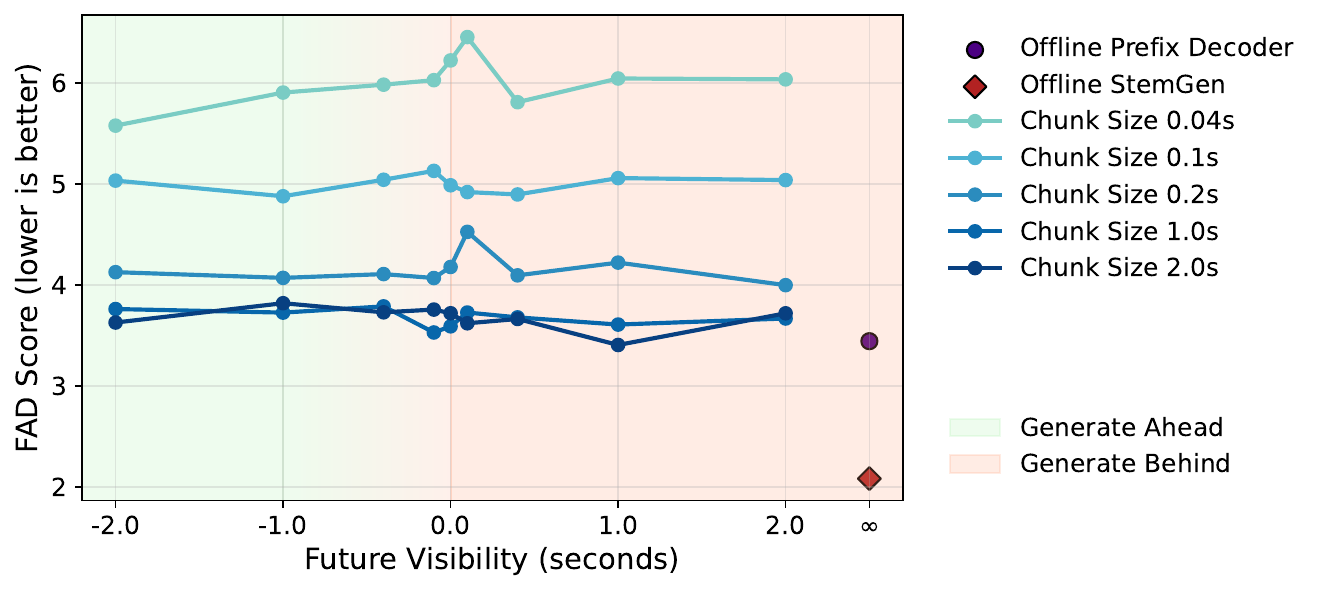}\label{fig:fad_score_chunk}}
  \caption{Beat alignment and FAD score of generated accompaniment for streaming models with combinations of $k>1$ and $t_f$.}
  \label{fig:metrics_appendix_beat_fad}
\end{figure*}

We provide a detailed breakdown of the evaluation metrics.
For streaming models with single-frame output ($k=1$), we report the decomposed COCOLA harmonic score and COCOLA percussive score across different future visibility $t_f$ in Fig.~\ref{fig:metrics_appendix}.
For streaming models with combinations of output chunk size $k>1$ and future visibility $t_f$, we present the results of other metrics: Fig.~\ref{fig:metrics_appendix_cocola_chunk} displays the COCOLA harmonic and percussive scores, while Fig.~\ref{fig:metrics_appendix_beat_fad} reports the rhythmic alignment (Beat $F_1$) and Audio Quality (FAD) scores.
Table~\ref{table:cocola_k1} lists the COCOLA scores plotted in Fig.~\ref{fig:COCOLA}.

\begin{table}[h]
\centering
\footnotesize
\setlength{\tabcolsep}{4pt}
\begin{tabular}{lccccccccccccc}
\hline
$t_f$ (s) & $-4$ & $-2$ & $-1$ & $-0.4$ & $-0.2$ & $-0.1$ & $0$ & $0.1$ & $0.2$ & $0.4$ & $1$ & $2$ & $4$ \\
\hline
COCOLA & 50.00 & 49.95 & 50.25 & 50.15 & 50.46 & 50.78 & 52.26 & 55.02 & 55.41 & 55.74 & 55.60 & 56.01 & 56.62 \\
\hline
\end{tabular}

\vspace{2mm}
\begin{tabular}{lcccc}
\hline
 & Offline StemGen & Offline Prefix Decoder & Ground Truth & Random Pairing \\
\hline
COCOLA & 60.41 & 55.68 & 61.27 & 47.94 \\
\hline
\end{tabular}
\caption{COCOLA scores plotted in Fig.~\ref{fig:COCOLA}: streaming models with $k=1$ across $t_f$ (in seconds) and the reference systems, with paired input and no prompt.}
\label{table:cocola_k1}
\end{table}

\newpage

\subsection{Results Under the Full-Mixture Setting}
\label{sec:fullmix}

\begin{figure*}[htbp]
  \centering
  \subfloat[COCOLA Score]{%
    \includegraphics[width=0.49\textwidth]{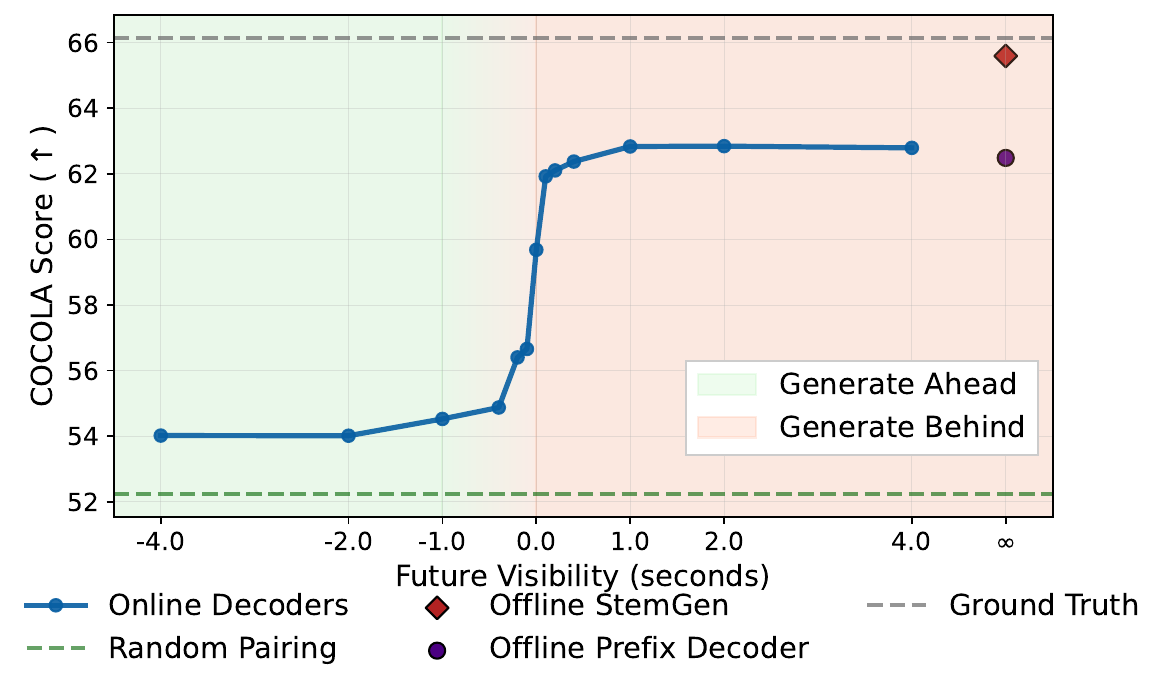}
    \label{fig:fullmix_cocola}}\hfill
  \subfloat[Rhythm Alignment]{%
    \includegraphics[width=0.49\textwidth]{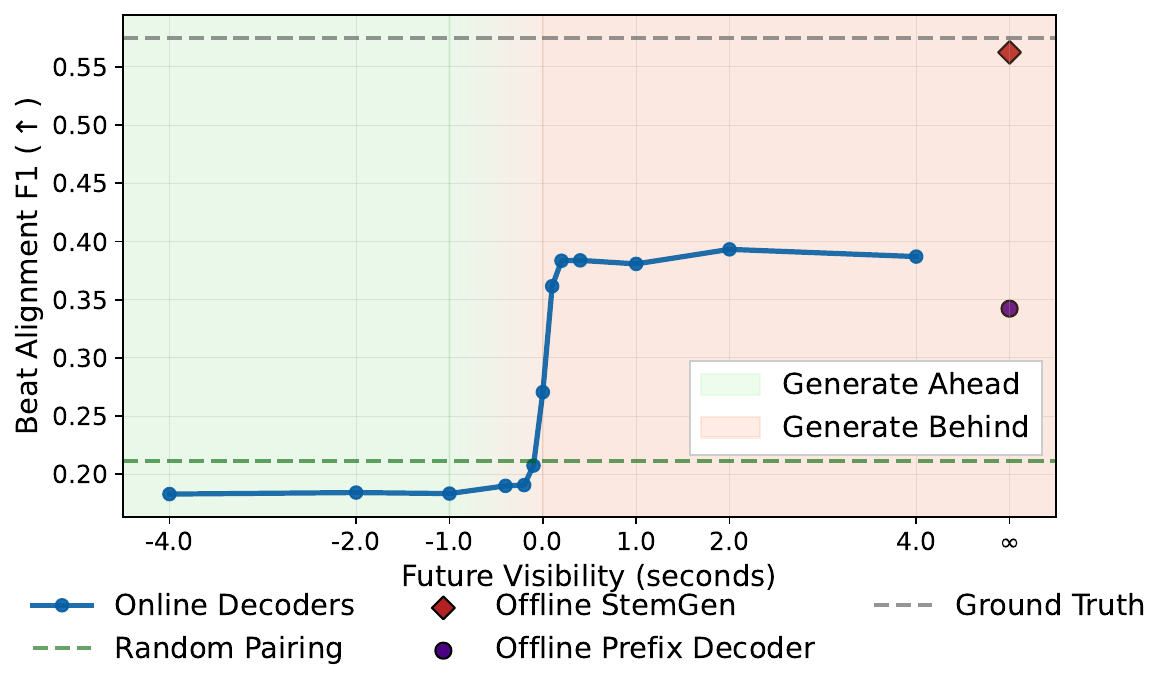}
    \label{fig:fullmix_beat}}\\
  \subfloat[Audio Quality]{%
    \includegraphics[width=0.49\textwidth]{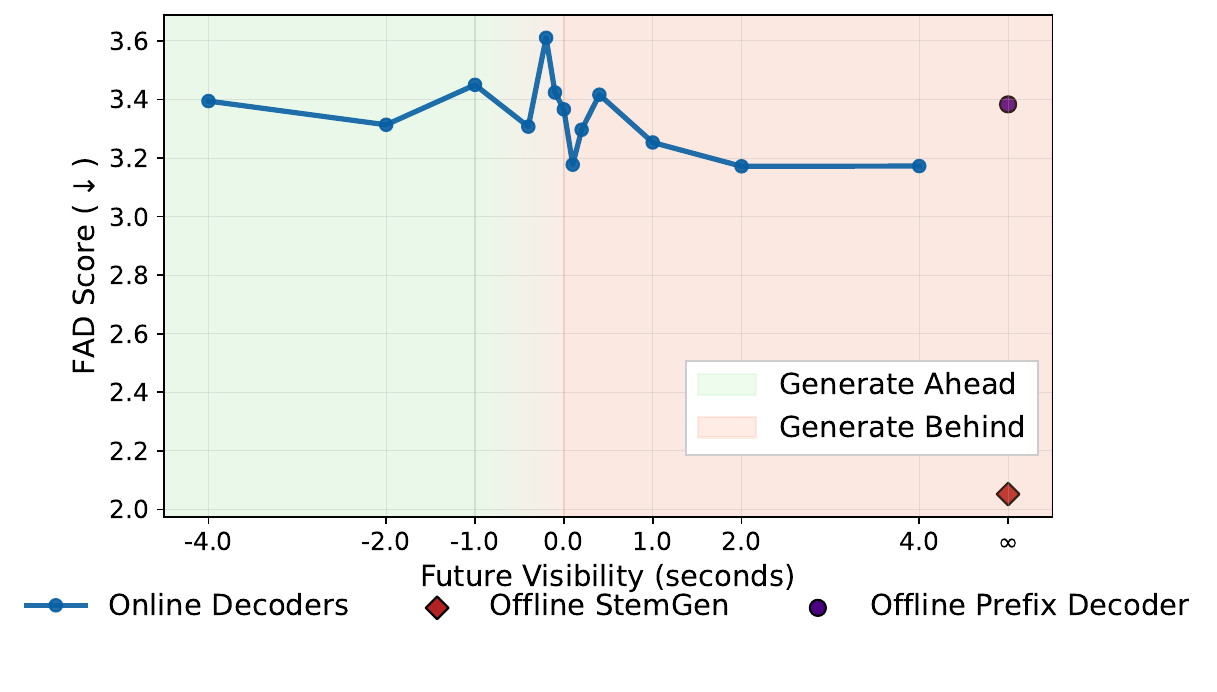}
    \label{fig:fullmix_fad}}
  \caption{COCOLA score, rhythm alignment and audio quality of generated accompaniment for streaming models with $k=1$ across different $t_f$ under the full-mixture setting.}
  \label{fig:fullmix_metrics}
\end{figure*}

\begin{figure*}[htbp]
  \centering
  \subfloat[COCOLA Score Across Prompt Length]{%
    \includegraphics[width=0.49\textwidth]{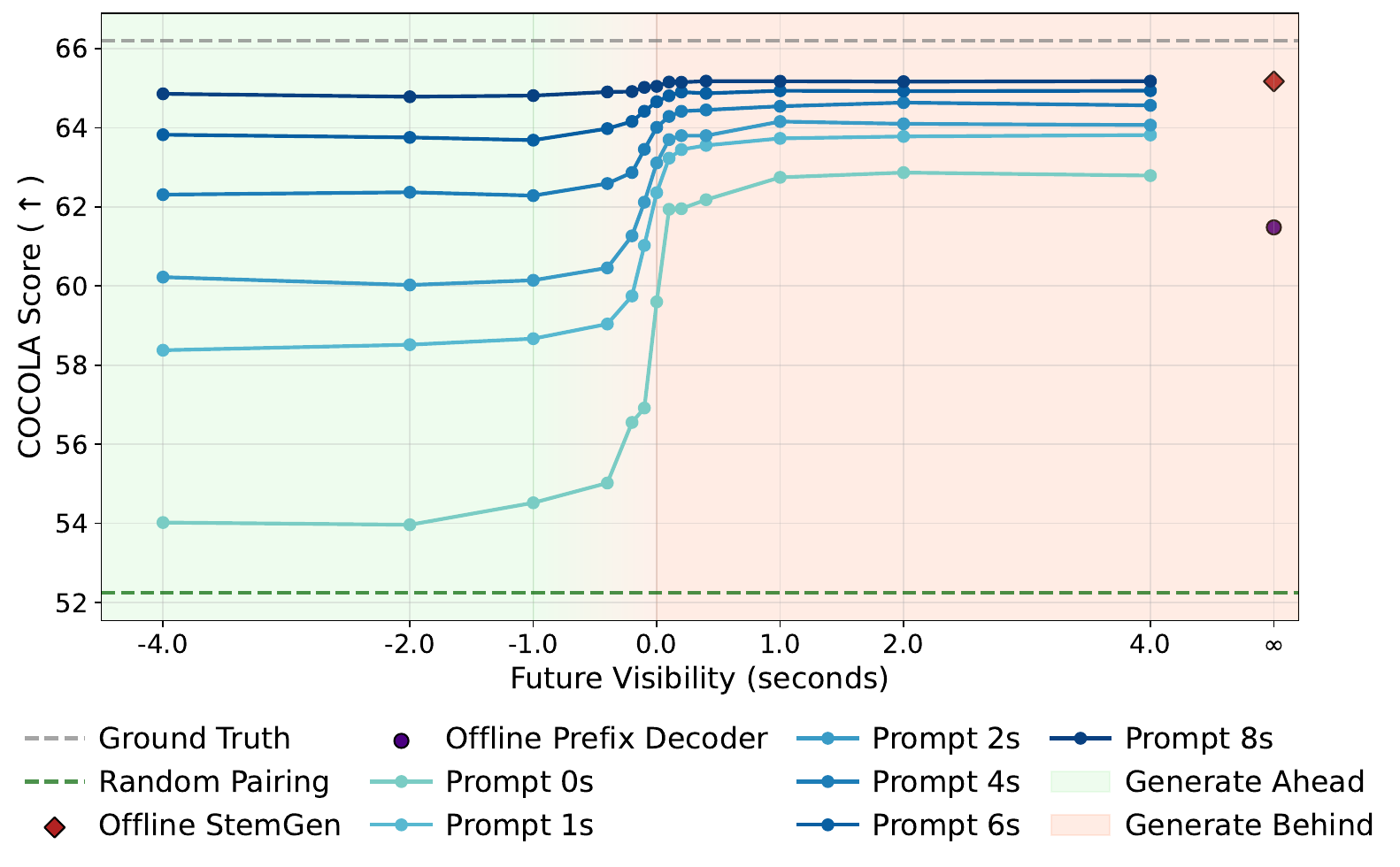}
    \label{fig:fullmix_prompt_length}}\hfill
  \subfloat[COCOLA Score Compared with Random Input]{%
    \includegraphics[width=0.49\textwidth]{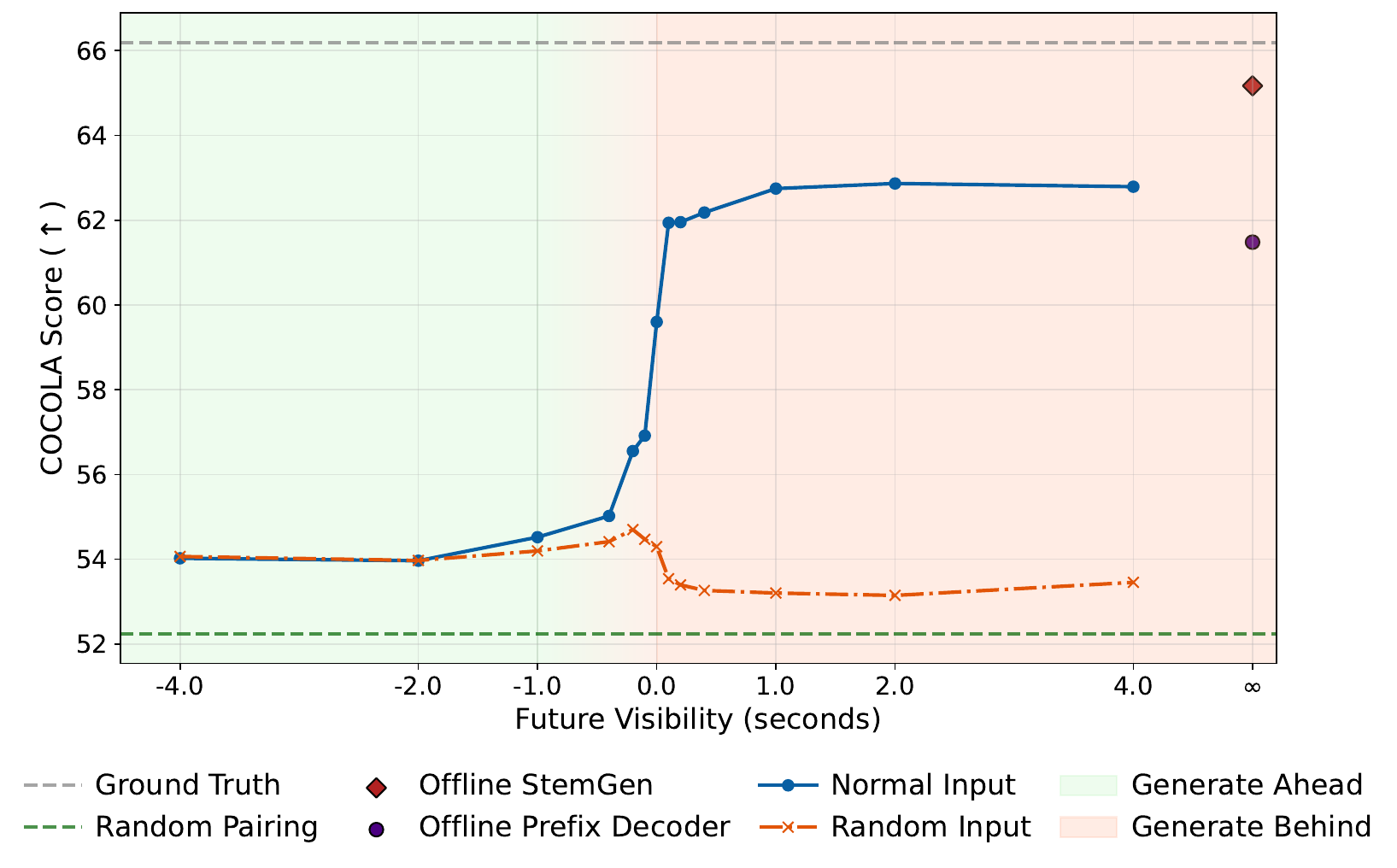}
    \label{fig:fullmix_random_input}}
  \caption{Accompaniment performance for streaming models with $k=1$ across different $t_f$ under the full-mixture setting. Left: performance under different prompt length. Right: performance when conditioning on the paired input compared with conditioning on a random input.}
  \label{fig:fullmix_prompt}
\end{figure*}

\begin{figure*}[htbp]
  \centering
  \includegraphics[width=0.6\textwidth]{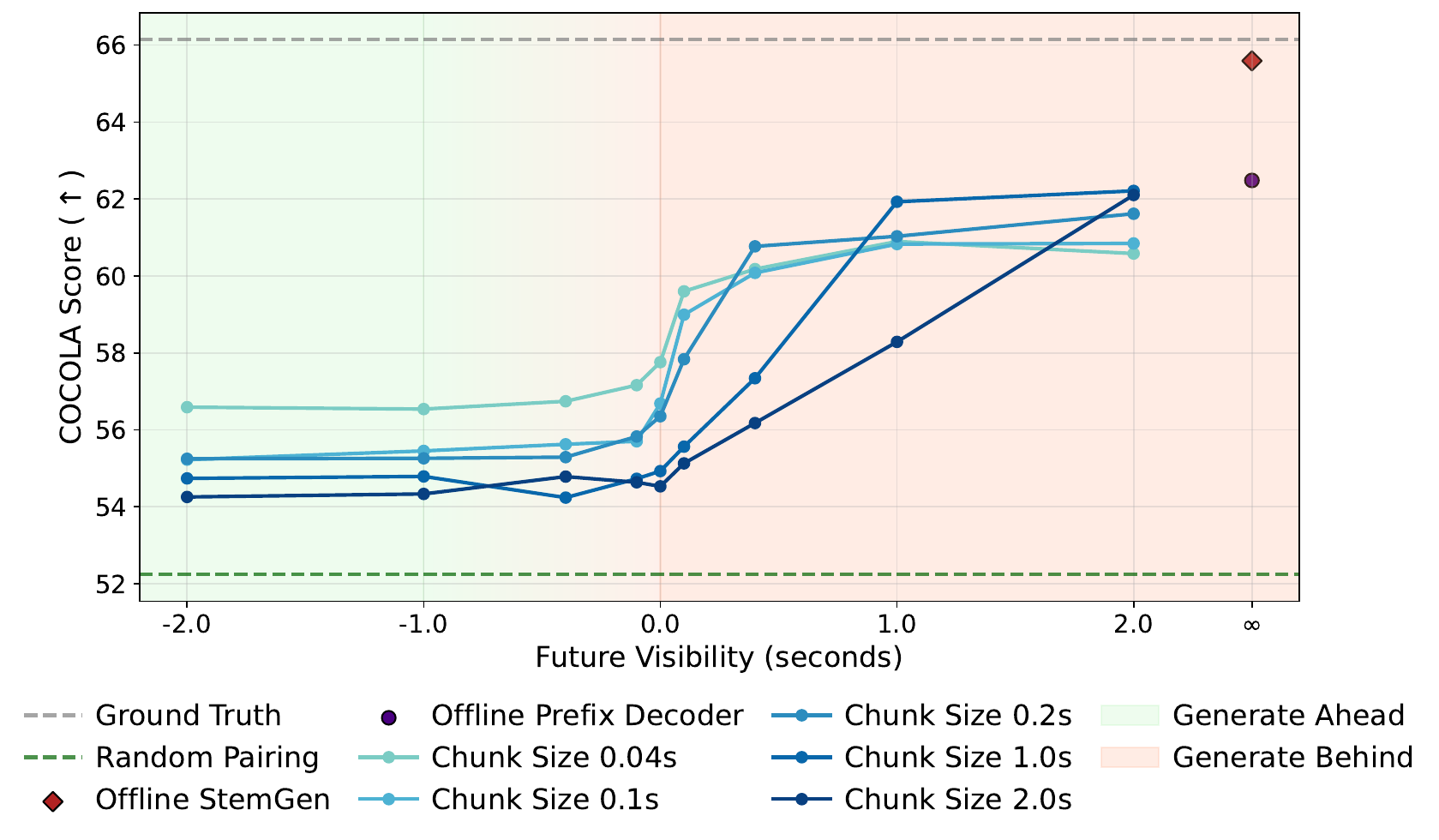}
  \caption{COCOLA score across $k>1$ and $t_f$ under the full-mixture setting.}
  \label{fig:fullmix_chunk}
\end{figure*}

An earlier version of our data pipeline contained a bug in input-mixture sampling: instead of drawing $N_{\text{in}}\sim\mathrm{Uniform}\{1,\dots,N-1\}$, every example used all $N-1$ remaining stems as the input mixture. In addition, the pre-computed $20\,\mathrm{s}$ windows of this setting contained only $10\,\mathrm{s}$ of audio followed by silence, so for streaming models with $t_f>0$ the last $t_f$ seconds of input lookahead were silent in both training and evaluation. The main text reports results in the corrected random-subset setting. Here we report the full set of experiments under the earlier full-mixture setting (Figs.~\ref{fig:fullmix_metrics}--\ref{fig:fullmix_chunk}), which constitutes a second experimental condition in which the model always hears the full mixture of accompanying stems. Because the test inputs differ, absolute scores are not comparable across the two settings (ground-truth COCOLA $66.1$ vs.\ $61.3$; random pairing $52.2$ vs.\ $47.9$), so we compare them relative to these reference levels.

The overall trends agree with the main results. Quality follows the same S-shaped dependence on $t_f$. Models with $t_f\le 0$ stay close to the random-pairing floor and largely ignore the input. Longer prompts help most when $t_f\le 0$. Chunks longer than the visible lookahead degrade coherence. Streaming models with sufficient lookahead match the offline prefix decoder, and StemGen performs best. We observe the following differences.
(i) Relative to the ground-truth and random-pairing levels, decoder models with $t_f\ge 0$ score higher in the full-mixture setting: for $k{=}1$ and $t_f{=}2\,\mathrm{s}$, the normalized COCOLA is $0.76$ vs.\ $0.61$, and paired inputs exceed random inputs by $9.7$ vs.\ $6.3$ points, whereas StemGen is similar in both settings ($0.96$ vs.\ $0.94$). The full-mixture setting is an easier task: with all other tracks present, the input carries more harmonic and rhythmic information about the target.
(ii) The transition around $t_f=0$ is wider: coherence already rises at $t_f=-0.2$ and $-0.1\,\mathrm{s}$ (normalized $0.30$ and $0.32$ vs.\ $0.19$ and $0.21$), and paired and random inputs already differ at $t_f=-0.2\,\mathrm{s}$ ($1.7$ vs.\ $0.2$ points).
(iii) FAD shows a mild degradation peak near $t_f=0$ and improves as $t_f$ moves away from zero in either direction, whereas in the random-subset setting it is nearly flat.
(iv) For $k>1$ and $t_f>0$, larger chunks perform better (at $t_f{=}2\,\mathrm{s}$, $k{=}1\,\mathrm{s}$ reaches $0.72$ vs.\ $0.60$ for $k{=}0.04\,\mathrm{s}$), and very small chunks ($0.04$--$0.2\,\mathrm{s}$) underperform, likely because each training example provides only $k$ frames of supervised targets. In the random-subset setting, chunk size matters little once $t_f\ge k$.
Here, normalized scores are $(s-s_{\text{rand}})/(s_{\text{GT}}-s_{\text{rand}})$, where $s_{\text{rand}}$ and $s_{\text{GT}}$ are the random-pairing and ground-truth scores of the respective setting.
Table~\ref{table:cocola_k1_fullmix} lists the COCOLA scores plotted in Fig.~\ref{fig:fullmix_cocola}, for comparison with Table~\ref{table:cocola_k1}.

\begin{table}[h]
\centering
\footnotesize
\setlength{\tabcolsep}{4pt}
\begin{tabular}{lccccccccccccc}
\hline
$t_f$ (s) & $-4$ & $-2$ & $-1$ & $-0.4$ & $-0.2$ & $-0.1$ & $0$ & $0.1$ & $0.2$ & $0.4$ & $1$ & $2$ & $4$ \\
\hline
COCOLA & 54.02 & 54.02 & 54.53 & 54.88 & 56.40 & 56.66 & 59.68 & 61.92 & 62.10 & 62.37 & 62.83 & 62.84 & 62.79 \\
\hline
\end{tabular}

\vspace{2mm}
\begin{tabular}{lcccc}
\hline
 & Offline StemGen & Offline Prefix Decoder & Ground Truth & Random Pairing \\
\hline
COCOLA & 65.59 & 62.48 & 66.14 & 52.24 \\
\hline
\end{tabular}
\caption{COCOLA scores plotted in Fig.~\ref{fig:fullmix_cocola}: streaming models with $k=1$ across $t_f$ (in seconds) and the reference systems under the full-mixture setting, with paired input and no prompt.}
\label{table:cocola_k1_fullmix}
\end{table}

\subsection{Example Lists}
We release the example lists that define every training, validation, and test example (track, target stem, input stems, and crop position), together with their SHA-256 hashes, so that the same examples, with identical audio, can be regenerated: \url{https://github.com/lukewys/stream-music-gen/tree/main/example_lists}.

\subsection{Transformer Backbone}
We use a transformer backbone similar to Llama 3. Following current large-model practice, the network uses pre layer normalization throughout \cite{xiong2020layernorm}, with Root Mean Square Normalization as the normalization operator \cite{zhang2019rmsnorm}; we adopt the simplified RMSNorm variant reported to work well at large scale \cite{qin2023transnormerllm}. Positional information is injected by rotary position embedding in self-attention \cite{su2021roformer}. Inside attention, we apply query-key normalization by normalizing queries and keys along the head dimension before the similarity computation \cite{henry2020qknorm}, together with a dimension-dependent scaling factor on the normalized scores as recommended by recent scaling results \cite{dehghani2023vit22b}. The feed-forward sublayers use the gated SwiGLU activation \cite{shazeer2020glu}. To improve decoding efficiency while preserving quality, key-value projections are shared across groups of query heads, i.e., grouped-query attention \cite{ainslie2023gqa}. The decoder also supports cross-attention for conditioning on external context.

\subsection{Listening Test Details}

The listening study (Fig.~\ref{fig:listening}) was conducted with models trained under the full-mixture setting (Appendix~\ref{sec:fullmix}). To ease cognitive load, we selected examples with at most six input tracks. In this study, the streaming model with $t_f=-1\,\mathrm{s}$ received fewer preference wins than even random pairing (Fig.~\ref{fig:listening}), although this difference is not significant after Bonferroni correction (Table~\ref{table:significance}).

\begin{table}[h]
\centering
\footnotesize
\begin{tabular}{lccccccc}
\hline
 & Ground Truth & Offline StemGen & Offline Prefix Decoder & $t_f = 1$ & $t_f = 0$ & $t_f = -1$ & Random Pairing \\
\hline
Ground Truth            & N/A & !   & *   & *   & *   & *   & * \\
Offline StemGen         & !   & N/A & !   & !   & !   & *   & * \\
Offline Prefix Decoder  & *   & !   & N/A & !   & !   & *   & ! \\
$t_f = 1$               & *   & !   & !   & N/A & !   & *   & ! \\
$t_f = 0$               & *   & !   & !   & !   & N/A & *   & ! \\
$t_f = -1$              & *   & *   & *   & *   & *   & N/A & ! \\
Random Pairing          & *   & *   & !   & !   & !   & !   & N/A \\
\hline
\end{tabular}
\caption{Pairwise statistical significance results for listening study. * indicates significant difference ($p < 0.05/21$), ! indicates non-significant ($p > 0.05/21$).}
\label{table:significance}
\end{table}

\end{document}